# 2ACT: AI-Accentuated Career Transitions via Skill Bridges

Drake Mullens
Tarleton State University, Stephenville, TX

*Stella Shen
University of Texas at Tyler, Tyler TX

*Corresponding author: fshen@patriots.uttyler.edu

**Abstract:**

This study introduces the AI-Accentuated Career Transitions framework, advancing beyond binary automation narratives to examine how distinct AI usage patterns reshape occupational mobility. Analyzing 545 occupations through multivariate modeling, we identify six qualitatively distinct human-AI usage patterns that differentially predict placement across job preparation zones. Our findings empirically validate the "missing middle" hypothesis: automation-focused usage strongly predicts lower job zone placement while augmentative usage predicts higher zones. Most significantly, we identify specific Knowledge, Skill, and Abilities combinations with AI usage patterns that function as "skill bridges" facilitating upward mobility. The interaction between task iteration AI usage and cognitive skills emerges as the strongest advancement predictor, creating pathways across traditionally disconnected occupational categories. Counterintuitively, despite directive AI's negative main effect, its interaction with technical knowledge positively predicts advancement in specialized domains. Comparative model testing confirms that AI usage patterns represent a distinct dimension of occupational classification that adds significant explanatory power beyond traditional skill measures. These findings reveal AI as a skill amplifier that widens capability gaps rather than an equalizing force. The 2ACT framework provides strategic guidance for workers, curriculum designers, policymakers, and organizations navigating increasingly AI-mediated career pathways.

**Keywords:** career mobility; artificial intelligence; skill bridges; human-ai collaboration; job zones

**Highlights**

- AI usage patterns explain occupational stratification beyond KSA measures, improving model pseudo-$R^2$ by 12%.
- Augmentative AI usage predicts higher job zones, while automation-focused AI usage predicts lower zones.
- AI amplifies existing skill advantages; workers with strong cognitive skills significantly benefit from advanced AI collaboration for promotion.
- Task iteration AI combined with cognitive skills creates the most powerful skill bridge for upward career mobility.
- A counterintuitive skill bridge: Directive AI usage, despite general negative effects, benefits workers with technical expertise in engineering fields.

- 

## 1. Introduction

Beyond alarmist narratives of AI-driven job displacement, AI is forging unprecedented career advancement pathways. The nature and mechanisms of these emerging pathways, however, remain largely unexplored. While contemporary research has evolved to assess AI at a granular task-level that recognizes occupational heterogeneity (Handa et al., 2025), attention has largely overlooked how AI creates new channels for career mobility (Agrawal et al., 2019, 2022). This preoccupation with displacement has obscured how AI already reshapes occupational boundaries and opens new career-mobility channels. AI-accentuated career transitions (2ACT) are restructuring occupational boundaries and creating novel pathways for career advancement that traditional models fail to capture.

Traditional models of career mobility have emphasized formal educational credentials as the primary mechanism for occupational advancement, but AI-driven task transformation may be creating alternative advancement pathways based on skill transferability and complementarity with AI systems. Recent empirical evidence from Anthropic (Handa et al., 2025) furthers this perspective by demonstrating that AI usage extends across approximately 36% of occupations for at least a quarter of their associated tasks with usage patterns concentrated in software development, technical writing, and analytical roles rather than comprehensively automating entire occupations. Identifying these emergent skill bridges between occupational categories is important for advancing evidence-based workforce development and reskilling policy.

The concept of job zones, representing levels of occupational preparation ranging from minimal (zone 1) to extensive (zone 5), provides a well-established framework for conceptualizing vertical occupational mobility (Peterson et al., 2001; National Center for O*NET Development, 2025). Research has consistently demonstrated that job zone advancement is associated with substantial wage premiums and improved employment stability (Modestino et al., 2020; Rothwell, 2015). However, the determinants of job zone placement in an AI-transformed economy remain insufficiently understood, particularly how distinct forms of human-AI collaboration might facilitate or constrain advancement.

We address this gap by developing and testing the 2ACT framework which differentiates between multiple forms of AI usage patterns: directive (where AI systems perform tasks with minimal human involvement), feedback loop (where AI receives environmental feedback to complete tasks), validation (where AI validates human outputs), task iteration (where humans and AI collaboratively refine outputs), learning (where AI provides knowledge and explanations), and thinking fraction (where users engage extended cognitive processes, a cross-cutting dimension that measures depth of cognitive engagement independent of whether the interaction is automative or augmentative). This multidimensional conceptualization builds empirical evidence from Anthropic's analysis of millions of Claude conversations which revealed that 57% of AI interactions demonstrate augmentative patterns while 43% show automation-focused usage (Handa et al., 2025). By distinguishing between these usage patterns, the 2ACT framework moves beyond simplistic automation narratives to capture how AI systems transform occupational tasks, potentially creating new forms of occupational connectivity that enable second acts in workers' career journeys. Figure 1 illustrates these six AI usage patterns and illustrates how automation-focused patterns are hypothesized to be associated with lower job zones while augmentation-focused patterns are hypothesized to be associated with higher job zone placement.

**Figure 1: 2ACT Framework: AI-Accentuated Career Transitions**

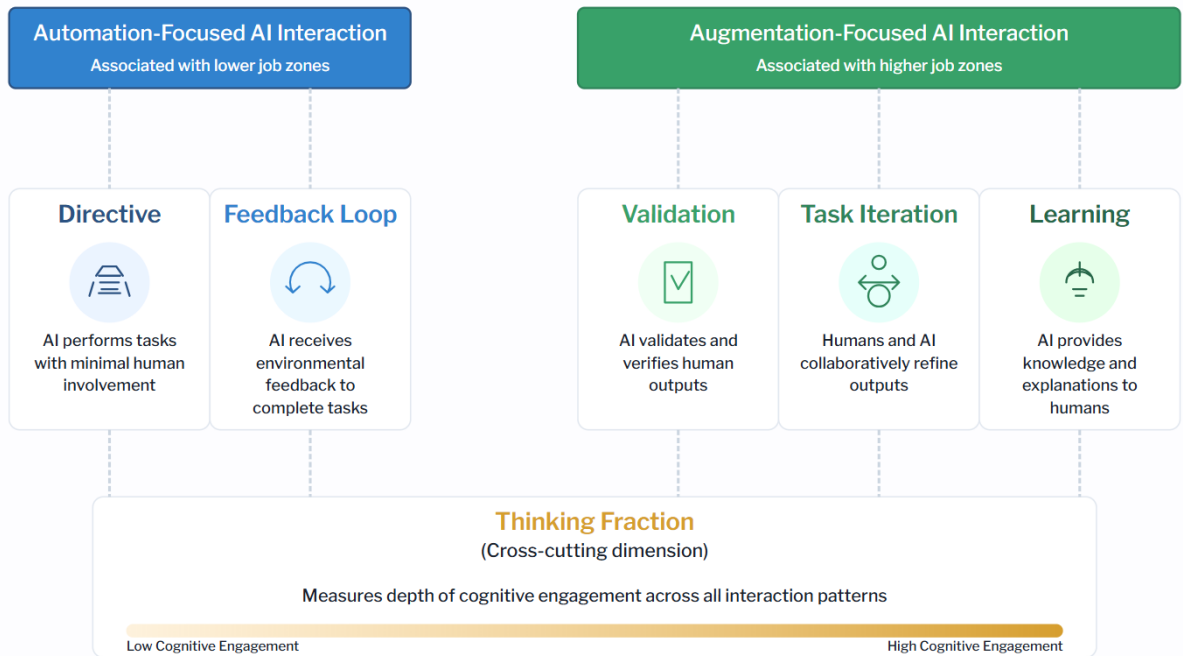

Note: The 2ACT Framework categorizes human-AI interactions into six distinct usage patterns. The Thinking Fraction is a cross-cutting dimension measuring cognitive engagement across all patterns.

Our central theoretical argument posits that specific AI usage patterns will differentially predict job zone placement, with directive and extended thinking metrics showing particular discriminatory power across the occupational hierarchy. Our analysis addresses three interconnected research questions:

1. How do AI usage patterns relate to job zone placement, and which patterns most strongly differentiate between job zone levels?

2. Which Knowledge, Skill, and Ability (KSA) components most strongly predict job zone placement when controlling for AI usage patterns?

3. What specific combinations of KSA components and AI usage patterns create the strongest predictors of higher job zone placement, potentially indicating effective skill bridges for career advancement?

This study's methodological innovation lies in its multivariate approach to modeling job zone classification. We integrate O*NET data on occupational KSA importance ratings with novel metrics quantifying six distinct patterns of human-AI usage, derived from task-level data and aggregated to the occupation level using task importance weights. We then employ principal component analysis, least absolute shrinkage and selection operator (LASSO) regression for feature selection, and ordinal logistic regression as our primary method to model job zone placement. The robustness of our findings is further assessed using machine learning (Random Forest) models, which also demonstrate strong predictive performance. This sequence allows us to critically examine the statistical interactions between KSA dimensions and AI usage patterns to identify potential skill bridges for career advancement.

By mapping how AI-induced task transformation, reflected in varying human-AI usage patterns, relates to occupational classification and career advancement opportunities, this study makes a timely contribution to understanding the evolving relationship between technological change and labor market dynamics. Our findings have significant practical implications for workers seeking strategic advancement, educational institutions designing curricula, policymakers developing workforce initiatives, and organizations implementing AI systems. As AI systems become increasingly pervasive, identifying the specific mechanisms through which they reshape career mobility pathways will be essential for developing effective strategies to navigate an AI-transformed economy.

## 2. Literature Review

The integration of artificial intelligence into workplace contexts fundamentally alters how we conceptualize occupational boundaries and career advancement. While research has extensively documented AI's disruptive potential (Brynjolfsson & McAfee, 2014; Frank et al., 2019), scholarly attention has predominantly focused on displacement effects rather than transformation opportunities (Frey & Osborne, 2017; Acemoglu & Restrepo, 2020). This emphasis on substitution has obscured how career pathways are facilitated by 2ACT.

### 2.1. AI Impact on Occupational Tasks

The analytical lens through which we examine technological impact has evolved significantly, moving from job-level automation predictions (Frey & Osborne, 2017) to granular task-based approaches (Autor, 2015; Arntz et al., 2017; Vendramin et al., 2021). This evolution reflects growing recognition that occupations comprise multiple tasks with varying susceptibility to technological change (Autor et al., 2003; Autor & Salomons, 2018). However, prior research has predominantly treated AI as a monolithic force rather than differentiating qualitatively distinct AI usage patterns. Davenport and Kirby (2016) began addressing this limitation by identifying five patterns of human-AI interaction (stepping up, stepping aside, stepping in, stepping narrowly, and stepping forward), while Vaccaro et al. (2024) provided evidence that different AI implementation approaches yield varied labor market outcomes. Our 2ACT framework extends these contributions by systematically differentiating six AI usage patterns and examining their relationship to occupational advancement.

Current empirical findings challenge the canonical skill-biased technological change hypothesis which predicts uniformly positive relationships between advanced technologies and higher skill requirements (Katz & Murphy, 1992; Berman et al., 1998). Instead, they align with task-biased frameworks (Autor et al., 2003; Acemoglu & Autor, 2011) suggesting differentiated impacts across the occupational spectrum. This demonstrates what Daugherty and Wilson (2018) termed the missing middle phenomenon. Handa et al.'s (2025) observation that AI usage peaks in mid-to-high wage occupations but drops at both extremes provides compelling empirical support for this differentiated impact, yet the specific mechanisms creating or constraining advancement opportunities remain underexplored.

### 2.2. Career Mobility and Skill Bridges

Traditional career mobility frameworks have emphasized formal educational credentials as the primary advancement mechanism (Holzer, 2015; Autor, 2019). However, network-based approaches to occupational mobility have revealed alternative pathways based on skill similarity (Gathmann & Schönberg, 2010; Neffke et al., 2017). This line of research conceptualizes the labor market as an interconnected system where proximity indicates mobility potential with skill bridges connecting different occupational clusters (Alabdulkareem et al., 2018).

Additionally, another stream of work has begun exploring how technological change reshapes these occupational networks. Muro et al. (2019) documented how digital technologies create new forms of occupational proximity while Frank et al. (2019) identified emerging patterns of complementary intelligence between human expertise and AI capabilities. These contributions suggest that AI may be creating new forms of occupational connectivity. However, a lacuna exists in systematically examining how specific AI usage patterns might facilitate or constrain movement across job zones.

Career theories, including Protean Career Theory (Hall, 2004; Haenggli et al., 2021) and boundaryless career concepts (Arthur & Rousseau, 2001; Sullivan & Arthur, 2006), emphasize non-linear progression and boundary-crossing mobility. These frameworks highlight adaptability and strategic skill development as essential resources for navigating complex career environments (Savickas, 2012). However, as noted by Hirschi (2018), these perspectives often lack concrete guidance on specific adaptive strategies in technology-transformed labor markets, another lacuna our research directly addresses by identifying specific skill bridges created through AI usage patterns.

### 2.3. Taxonomy of Collaboration Modes

Emerging research has documented distinct patterns of human-AI collaboration. Raisch and Krakowski (2021) identified varied collaboration models based on decision rights allocation, while Jarrahi (2018), Jarrahi et al. (2023), and Baker et al. (2024) cataloged emerging patterns of AI integration across knowledge-intensive domains. Handa et al.'s (2025) empirical findings (see sections 1 and 2.1) provide concrete evidence for these theoretical models.

Research on skill complementarity has similarly identified how technological change differentially affects skill domains. Deming (2017) documented increasing returns to social skills in the context of technological change, while Deming and Noray (2020) found that the half-life of technical skills is shortening. Felten et al. (2021) demonstrated that AI capabilities differentially complement or substitute for human skills across occupational categories. These contributions collectively suggest that particular skill-technology combinations may create distinct advancement opportunities, yet research addressing how AI usage patterns moderate relationships between skills and career mobility remains absent.

This lack of granular understanding significantly constrains our ability to provide actionable guidance for workers, educators, and policymakers operating in an AI-transformed labor markets. Without identifying which specific AI usage patterns facilitate advancement across job zones and which skill-AI combinations create effective bridges for upward mobility, stakeholders cannot develop strategies for career development.

### 2.4. Research Gaps and Opportunities

Despite substantive advances in understanding both AI's impact on work and career mobility patterns, several critical gaps remain that our research addresses:

First, existing frameworks treat AI impact predominantly through a binary lens of substitution versus complementarity, neglecting the qualitatively distinct AI usage patterns that may create differentiated career advancement opportunities. By distinguishing between these usage patterns, the 2ACT framework moves beyond simplistic automation narratives to capture how AI systems transform occupational tasks, potentially creating new forms of occupational connectivity that facilitate diverse career trajectories

Second, while occupational mobility networks have been extensively mapped, limited research has examined how technological change creates new mobility opportunities across job zones. Our analysis of how AI usage patterns relate to job zone classification addresses this gap, identifying specific pathways that facilitate upward mobility across traditionally disconnected occupational categories.

Third, though emerging research documents various forms of human-AI collaboration, the relationship between these collaboration patterns and career advancement remains undertheorized. By identifying which AI usage patterns predict higher job zone placement and which create barriers to advancement, our research provides essential insights into the career implications of different human-AI collaboration approaches.

Fourth, despite the wealth of theoretical perspectives on career development in the digital age, practical guidance for individuals immersed in an AI-transformed labor markets remains limited. By identifying specific skill bridges created through AI usage patterns, our research provides concrete strategies for career advancement in an AI-transformed economy.

In addressing these gaps, our study makes a significant theoretical contribution to understanding career mobility in the age of AI while providing actionable guidance for stakeholders navigating an increasingly AI-integrated labor market. Through the 2ACT framework, we move beyond simplistic automation narratives to examine how different AI usage patterns create qualitatively distinct career advancement opportunities and reveal previously invisible pathways across occupational boundaries.

## 3. Hypotheses

Our investigation into 2ACT is guided by hypotheses concerning the distinct roles of AI usage patterns, foundational worker KSAs, their synergistic combinations, and the overall explanatory contribution of AI usage patterns.

### 3.1. AI Usage Patterns and Job Zone Placement

The impact of AI on occupations is not monolithic; rather, it is contingent upon how AI is used in relation to human tasks. Automation-focused AI usage, which tends to substitute for human labor in routine, codifiable tasks, is often concentrated in lower and middle-skill occupations (Autor, 2015; Handa et al., 2025). Conversely, augmentative AI usage, which complements and enhances human capabilities, is more likely to be associated with complex tasks prevalent in higher-skill occupations (Daugherty & Wilson, 2018; Handa et al., 2025). Based on this task-biased technological change perspective, we hypothesize:

- **H1a:** Automation-focused AI usage patterns (e.g., Directive and Feedback Loop) will be negatively associated with job zone placement.

- **H1b:** Augmentation-focused AI usage patterns (e.g., Task Iteration) will be positively associated with job zone placement.

### 3.2. KSA Dimensions and Job Zone Placement

Job zone placement, reflecting occupational preparation levels, is fundamentally linked to an occupation's KSA requirements. Skill-biased technological change literature has consistently documented increasing labor market returns to cognitive and analytical skills (Deming, 2017; Liu & Grusky, 2013). Conversely, skills associated with routine manual tasks or those more susceptible to automation may see declining relative demand or value (Autor et al., 2003). Furthermore, the type of knowledge required can differentiate occupational tiers, with knowledge related to complex human and social systems often commanding a premium over more routinized production knowledge (Liu & Grusky, 2013). We therefore propose:

- **H2a:** Cognitive/analytical skill dimensions will be positively associated with job zone placement.
- **H2b:** Skill dimensions primarily representing resource management and practical tasks will be negatively associated with job zone placement.
- **H2c:** Knowledge dimensions emphasizing social/human sciences will be positively associated with job zone placement, while those emphasizing physical/engineering production knowledge) will be negatively associated with job zone placement.

### 3.3. Statistical Interaction Effects: Skill-AI Usage Combinations as Career Bridges and Barriers

The transformative potential of AI likely emerges not just from its direct effects, but from its interplay with human KSAs. Effective career advancement in an AI-integrated economy may depend on workers' ability to leverage AI in ways that complement their existing skills, creating synergistic "skill bridges" (Brynjolfsson & Mitchell, 2017). For instance, advanced cognitive skills combined with iterative AI collaboration may unlock significant value. Conversely, certain skill sets might become particularly vulnerable or create "career barriers" when combined with specific AI usage patterns, especially those automating coordination or oversight functions (Tolan et al., 2021). Similarly, the capacity for humans to validate AI outputs or "step in" can be crucial in specialized domains (Davenport & Kirby, 2016). Thus, we hypothesize:

- **H3a:** Task Iteration AI usage will positively moderate the relationship between cognitive/analytical skill dimensions and job zone placement, such that the positive association of cognitive/analytical skills is stronger when Task Iteration AI usage is high.
- **H3b:** Directive AI usage will negatively moderate the relationship between resource management skill dimensions and job zone placement, such that the negative association of resource management skills is stronger (or any positive association is weaker) when Directive AI usage is high.
- **H3c:** Validation AI usage will positively moderate the relationship between natural/health sciences knowledge dimensions and job zone placement, such that the positive association of this knowledge is stronger when Validation AI usage is high.

### 3.4. Comparative Explanatory Power of AI Usage Patterns

If, as our 2ACT framework posits, distinct AI usage patterns constitute an emerging axis of occupational differentiation, then metrics capturing these patterns should provide explanatory power for job zone placement beyond that offered by traditional KSA measures alone. Establishing this would underscore the unique contribution of understanding human-AI collaboration modalities in analyzing contemporary labor market stratification. We therefore hypothesize:

- **H4:** AI usage metrics will contribute significant additional explanatory power to models predicting job zone placement, over and above the explanatory power provided by KSA dimensions alone.

## 4. Methods
## 4.1. Data Sources and Sample

Our analysis utilizes a composite dataset integrating two primary sources of occupation-level information:

1. **O*NET Database (Version 26.0)**: We extracted KSA importance ratings for 545 distinct occupations. These ratings reflect the importance of 120 distinct elements across knowledge domains (33 elements), skills (35 elements), and abilities (52 elements).

2. **AI Usage Metrics**: We incorporated six distinct AI usage metrics for each occupation (Directive, Feedback Loop, Validation, Task Iteration, Learning, and Thinking Fraction). These metrics were initially derived from the task-level classifications of AI usage published by Handa et al. (2025), which were mapped to O*NET task statements. To transform these task-level AI usage scores into occupation-level metrics for our analysis, we performed an aggregation for each occupation. Specifically, for each of the six AI usage patterns, we calculated a weighted average of its task-level score across all O*NET tasks constituting an occupation. The O*NET importance rating of each constituent task served as the weight in this calculation, ensuring that AI usage on more critical tasks had a greater influence on the occupation's overall AI usage profile. This methodological approach leveraged the established task classifications from Handa et al. (2025) while extending their work through novel occupation-level statistical modeling focused on career mobility patterns.

3. **Job Zone Classification**: Our dependent variable is the O*NET Job Zone classification (1-5), which categorizes occupations based on levels of preparation, education, training, and experience required.

The final analytical sample included 545 occupations with complete data across all variables, representing approximately 54% of the occupations in the O*NET database and including representatives from all major occupational categories in the U.S. labor market.

## 4.2. Analytical Approach and Implementation

Python 3.13 with relevant packages was used for robust data preparation and modeling. Missing data was handled using Multiple Imputation by Chained Equations (sklearn.impute.IterativeImputer). The relationship between features and job zone outcomes was modeled using both Proportional Odds Ordinal Logistic Regression (statsmodels.miscmodels.ordinal_model.OrderedModel) and Random Forest classification (sklearn.ensemble.RandomForestClassifier) and regressor (sklearn.ensemble.RandomForestRegressor). Model performance was assessed using stratified 5-fold cross-validation (sklearn.model_selection.StratifiedKFold).

## 4.3. Dimensionality Reduction and Feature Selection

Given the high dimensionality of the KSA variables, we employed Principal Component Analysis (PCA) as a dimensionality reduction technique (Hair et al., 2010). PCA was conducted separately for KSA domains to maintain domain-specific interpretability. Component loadings were retained based on the combined criteria of scree plot examination, a minimum threshold of 65% explained variance, and the interpretability of resulting components in relation to the underlying occupational factors being measured (Factor loadings for each component are provided in Appendix A):

1. **Knowledge Domain**: Three principal components were retained, explaining 68.4% of variance. 1. represents Social/Human Sciences vs. Physical/Engineering Production knowledge, 2. represents Engineering/Physical Sciences knowledge, and 3. represents Natural & Health Sciences vs. Business Operations knowledge.

2. **Skills Domain**: Three principal components were retained, explaining 72.1% of variance. 1. represents Cognitive/Analytical vs. Hands-On/Technical skills, 2. represents Technical/Diagnostic vs. Interpersonal/Service skills, and 3. represents Resource Management & Practical vs. Abstract/Scientific skills.

3. **Abilities Domain**: Three principal components were retained, explaining 65.7% of variance. 1. represents Physical/Psychomotor vs. Cognitive/Verbal abilities, 2. represents Perceptual Reasoning abilities, and 3. represents Creative/Expressive vs. Detail-Oriented abilities.

To identify the most relevant predictors of job zone placement from our initial set of 69 potential predictors (9 KSA principal components, 6 AI usage metrics, and 54 interaction terms) and address multicollinearity concerns, we employed the Least Absolute Shrinkage and Selection Operator (LASSO) regression technique (Hastie et al., 2015; Tibshirani, 1996). This process selected 36 variables with non-zero coefficients, indicating their relevance for predicting job zone placement. These selected variables were subsequently used in the final ordinal logistic regression model.

### 4.4. Primary Analysis: Ordinal Logistic Regression

Given the ordinal nature of our dependent variable (job zones 1-5), we employed ordinal logistic regression as our primary analytical approach. Following our feature selection process, we specified a model that included both main effects and interaction terms, implemented using the OrderedModel class from the statsmodels package in Python with the logit link function and maximum likelihood estimation.

The final model included:

1. Main effects for the PCA components selected by LASSO
2. Main effects for the AI usage metrics
3. Theoretically important interaction terms between KSA components and AI usage patterns

Model fit was assessed using McFadden's Pseudo R-squared, with a value of 0.539 indicating moderate predictive power. The proportional odds assumption was tested using a likelihood ratio test comparing the ordinal model against a non-restricted alternative. The non-significant result ($p > .05$) indicates the proportional odds assumption was not violated. To evaluate multicollinearity, we evaluated the VIF for each predictor in the model. All but two variables were under the 5 threshold; S_PC1 and A_PC1 showed elevated VIF values (approximately 9.8), indicating potential multicollinearity. While these values are below the commonly used threshold of 10 (Hair et al., 2010), they suggest the need for cautious interpretation of these coefficients.

### 4.5. Supplementary Analysis: Random Forest

To complement our primary analysis and explore potential non-linear relationships, we conducted supplementary analyses with machine learning using Random Forest models (Breiman, 2001). First, we implemented a Random Forest Classifier with 500 trees and balanced class weights, evaluated using 5-fold stratified cross-validation. This model achieved an overall accuracy of 0.6716 (SD = 0.0408) and a balanced accuracy of 0.5745 (SD = 0.0310), confirming that the selected features possess predictive power for classifying job zones. Feature importance analysis using Mean Decrease in Impurity (MDI) from this classifier identified cognitive/analytical skills (S_PC1, importance = 0.201) and physical/psychomotor abilities (A_PC1, importance = 0.144) as the strongest predictors, followed by knowledge components and other skills. Among AI usage metrics, task iteration (importance = 0.046) and directive interaction (importance = 0.040) showed the strongest relationships.

Separately, to generate additional metrics, we also fit a Random Forest Regressor treating job zone as a continuous variable. This specific analysis yielded an $R^2$ score of 0.793 on test data, a Mean Squared Error (MSE) of 0.283, and an Out-of-Bag (OOB) $R^2$ score of 0.787.

These supplementary analyses, using both classification and regression approaches with Random Forest, broadly support the importance of the key KSA dimensions and AI usage patterns identified in our primary ordinal logistic regression model.

### 4.6. Methodological Limitations

Several methodological limitations should be acknowledged:

- **First,** the cross-sectional nature of our data limits causal inference regarding the relationship between AI usage patterns and job zone progression; longitudinal or experimental designs would be necessary to establish causality definitively.

- **Second,** our occupation-level analysis may mask within-occupation heterogeneity in AI usage patterns.

- **Third,** the O*NET KSA importance ratings are based on self-reported data, which may introduce subjective biases. Furthermore, these ratings reflect occupations that continue to exist, potentially leading to a form of survivor bias where declining or obsolete roles are under-represented.

- **Fourth,** while we leverage Anthropic's comprehensive analysis of AI usage patterns, the mapping from task-level metrics (and the underlying classification of those AI interactions) to occupation-level metrics necessarily involves aggregation that may obscure nuanced patterns of AI usage or misclassify rare/emergent AI usage behaviors.

- **Finally,** our focus on the U.S. labor market potentially limits generalizability to countries with different occupational structures or non-English language contexts where AI usage patterns might differ.

Despite these limitations, our methodology represents a rigorous approach to examining the relationship between AI usage patterns and job zone placement and provides valuable insights into potential career mobility pathways facilitated by the 2ACT framework.

## 5. Results
### 5.1. Principal Component Structure and Model Overview

Principal components derived from our PCA of KSA domains captured meaningful skill dimensions. The Knowledge components distinguished between social/human sciences and physical/engineering production (K_PC1), specialized engineering and physical sciences knowledge (K_PC2), and natural/health sciences versus business operations knowledge (K_PC3). The Skills components differentiated cognitive/analytical from hands-on technical skills (S_PC1), technical/diagnostic from interpersonal/service skills (S_PC2), and resource management from abstract/scientific skills (S_PC3). The Abilities components contrasted physical/psychomotor with cognitive/verbal abilities (A_PC1), identified perceptual reasoning abilities (A_PC2), and distinguished creative/expressive from detail-oriented abilities (A_PC3).

Random Forest analysis revealed that cognitive/analytical skills (S_PC1, importance = 0.20) and physical/psychomotor abilities (A_PC1, importance = 0.14) were the strongest predictors of job zone placement, followed by engineering/physical sciences knowledge (K_PC2, importance = 0.01). Among AI usage metrics, task iteration showed the highest importance (0.05), followed by directive usage (0.04). The model demonstrated strong predictive performance with an $R^2$ of 0.79.

### 5.2. Direct Relationships with Job Zone Placement

Our ordinal logistic regression analysis yielded a model with good fit (McFadden's Pseudo $R^2$ = 0.54). Table 1 presents key coefficients from this analysis.

**Table 1. Selected Ordinal Logistic Regression Results Predicting Job Zone Placement**

| Variable | Coefficient | Std. Error | p-value |
|---|---|---|---|
| **Skill Dimensions** | | | |
| K_PC1 (Social/Human Sciences) | 0.36 | 0.07 | <0.001 |
| K_PC3 (Natural & Health Sciences) | 0.26 | 0.08 | 0.001 |
| S_PC1 (Cognitive/Analytical Skills) | 0.70 | 0.09 | <0.001 |
| S_PC2 (Technical/Diagnostic Skills) | 0.29 | 0.09 | 0.001 |

| | | | | |
|---|---|---|---|---|
| S_PC3 (Resource Management Skills) | -0.78 | 0.10 | <0.001 | |
| **AI Usage Patterns** | | | | |
| Directive | -2.58 | 1.01 | 0.01 | |
| Feedback Loop | -15.23 | 6.19 | 0.01 | |
| Task Iteration | 2.19 | 1.15 | 0.06 | |
| Thinking Fraction | 10.29 | 5.98 | 0.09 | |
| **Key Interaction Effects** | | | | **Interpretation** |
| Directive × K_PC2 | 1.03 | 0.41 | 0.01 | **Surprising positive effect**: Directive AI with engineering knowledge predicts higher job zones |
| Directive × S_PC3 | -1.41 | 0.60 | 0.02 | **Career barrier**: Directive AI with resource management skills strongly predicts lower job zones |
| Validation × K_PC3 | 8.10 | 3.73 | 0.03 | **Skill bridge**: Validation AI with scientific knowledge strongly predicts higher job zones |
| Task Iteration × S_PC1 | 1.10 | 0.29 | <0.001 | **Primary skill bridge**: Task iteration with cognitive skills strongly predicts higher job zones |
| **Additional Interaction Effects** | | | | |
| Directive × A_PC1 | 0.44 | 0.23 | 0.06 | Directive AI moderates negative effect of physical abilities (marginal) |
| Feedback Loop × A_PC1 | 2.25 | 1.35 | 0.10 | Feedback loop AI with physical abilities weakly predicts higher job zones (marginal) |
| Validation × A_PC2 | -3.50 | 1.93 | 0.07 | Validation AI with perceptual reasoning predicts lower job zones (marginal) |
| Task Iteration × S_PC2 | -0.87 | 0.46 | 0.06 | Task iteration with technical skills predicts lower job zones (marginal) |
| Learning × S_PC3 | -0.95 | 0.48 | 0.05 | Learning AI with resource management skills predicts lower job zones |

| Learning × A_PC3 | 0.74 | 0.38 | 0.05 | Learning AI with creative abilities predicts higher job zones (marginal) |

Note: Significance levels: * p < .05, ** p < .01, *** p < .001. Marginal significance (0.05 < p < 0.10) indicated where relevant. Model McFadden's Pseudo $R^2$ = 0.54. All predictors were centered prior to analyses to limit multicollinearity and facilitate interaction interpretation.

Our analysis provides strong support for H1a with both directive usage ($\beta$ = -2.58, p = .01) and feedback loop usage ($\beta$ = -15.23, p = .01) showing significant negative relationships with job zone placement. H1b receives partial support with task iteration showing a positive relationship approaching conventional significance ($\beta$ = 2.19, p = .06) and thinking fraction demonstrating a marginally significant positive association ($\beta$ = 10.29, p = .09).

Similarly, the modeled results demonstrate strong support for hypotheses regarding skill dimensions and job zone placement. Consistent with H2a, cognitive/analytical skills (S_PC1) show a strong positive association with job zone placement ($\beta$ = 0.70, p < .001), confirming these skills are highly rewarded in occupations requiring greater preparation. H2b is also strongly supported with resource management skills (S_PC3) showing a significant negative association ($\beta$ = -0.78, p < .001). The results provide support for H2c with social/human sciences knowledge (K_PC1) showing a significant positive association ($\beta$ = 0.36, p < .001). Figure 2 illustrates the varying average usage of different AI usage patterns across the five O*NET job zones.

**Figure 2: Average Prevalence of AI Usage Patterns Across O*NET Job Zones**

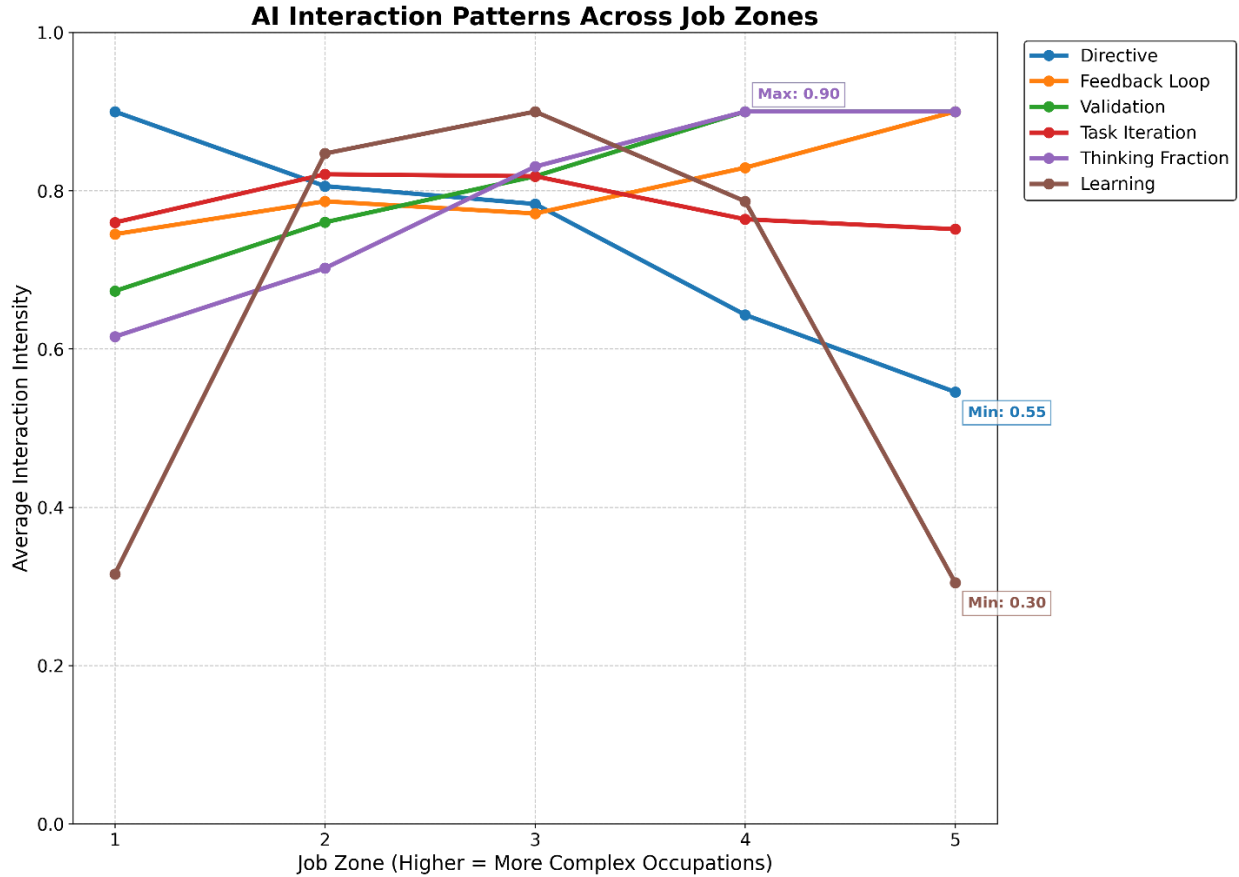

Note: Average AI usage pattern prevalence across job zones (1=minimal to 5=extensive preparation). Max/Min annotations identify highest/lowest values observed.

The stronger negative effects for automation-focused interactions compared to the positive effects for augmentative interactions suggest an asymmetric impact of AI on career mobility. While automation-focused AI usage strongly predicts lower job zone placement, the career-enhancing effects of augmentative interaction appear more modest.

### 5.3. Interaction Effects: Skill Bridges and Barriers

Beyond the main effects, the interplay between AI usage patterns and KSA dimensions reveals mechanisms of occupational mobility, illuminating specific skill bridges and barriers. Figure 3 visually articulates the most salient of these interactions, plotting predicted probabilities of higher (Job Zone ≥ 4) or lower (Job Zone ≤ 2) placement from our ordinal logistic regression model.

Our results confirm H3a, revealing a potent primary skill bridge: the interaction between task iteration AI usage and cognitive/analytical skills (S_PC1) demonstrates a significant positive effect ($\beta = 1.10$, $p < .001$). This multiplicative relationship, substantially exceeding the already significant main effect of cognitive skills, is interactions, detailed in Table 1 and visualized in Figure 3, demonstrates the interplay shaping occupational advancement in an AI-transformed economy.

H3a posited that Task Iteration AI usage would positively moderate the relationship between cognitive/analytical skills and job zone placement. Our results robustly confirm this hypothesis. The interaction between task iteration AI and cognitive/analytical skills (S_PC1) exhibits a significant positive effect ($\beta = 1.10$, $p < .001$). This interaction, substantially exceeding the already significant positive main effect of cognitive skills, indicates that when workers endowed with strong cognitive capabilities engage in iterative refinement processes with AI systems, this combination is a particularly potent predictor of higher job zone placement. As illustrated in Figure 3 (Panel A), the probability of an occupation being classified in Job Zone $\geq 4$ demonstrably increases with the prevalence of task iteration AI usage; however, this positive gradient is markedly steeper for individuals possessing high cognitive/analytical skills. This relationship suggests that iterative AI collaboration serves as a primary skill bridge, facilitating upward mobility, especially within cognitively-intensive occupational strata.

H3b, which anticipated that Directive AI usage would negatively moderate the relationship between resource management skills and job zone placement, also receives strong support. The interaction between directive AI and resource management skills (S_PC3) is significant and negative ($\beta = -1.41$, $p = .02$). Figure 3 (Panel B) visually articulates this dynamic, showing that increasing directive AI usage, when coupled with high resource management skills, markedly elevates the probability of an occupation being classified in Job Zone $\leq 2$. This finding suggests that this particular skill-AI combination functions as a career barrier, where the automation of tasks traditionally associated with resource management constrains advancement opportunities for individuals whose primary skill set aligns with these activities.

Furthermore, we find compelling support for H3c, which hypothesized a positive moderation by Validation AI usage on the relationship between natural/health sciences knowledge and job zone placement. The interaction between validation AI and natural/health sciences knowledge (K_PC3) is statistically significant and positive ($\beta = 8.10$, $p = .03$). This substantial interaction effect, depicted in Figure 3 (Panel C), reveals a steep increase in the probability of higher job zone placement (Job Zone $\geq 4$) when high levels of scientific knowledge are combined with validation-oriented AI interactions. This pattern suggests that in scientific and healthcare domains, AI systems that validate, rather than replace, human expert work create valuable complementarities. This forms an effective professional skill bridge, potentially facilitating vertical mobility in fields such as biomedical research and advanced healthcare diagnostics where expert oversight and AI-assisted verification are paramount.

Beyond our hypothesized interactions, a counterintuitive yet significant positive interaction emerges between directive AI usage and engineering/physical sciences knowledge (K_PC2, $\beta = 1.03$, $p = .01$). Despite the overall negative main effect of directive AI usage ($\beta = -2.58$, $p = .01$), its combination with strong technical knowledge predicts higher job zone placement. Figure 3 (Panel D) illustrates this surprising finding: in contexts characterized by high engineering/physical sciences knowledge, increasing directive AI usage is associated with a greater probability of placement in Job Zone $\geq 4$. This suggests that within highly specialized technical fields, the ability to effectively delegate routine, codifiable tasks to AI systems (directive usage) may not supplant but rather complement deep human expertise by freeing capacity for higher-order engineering, design, or analytical activities, thereby creating an unexpected advancement pathway.

**Figure 3: AI-Skill Interactions as Career Bridges and Barriers**

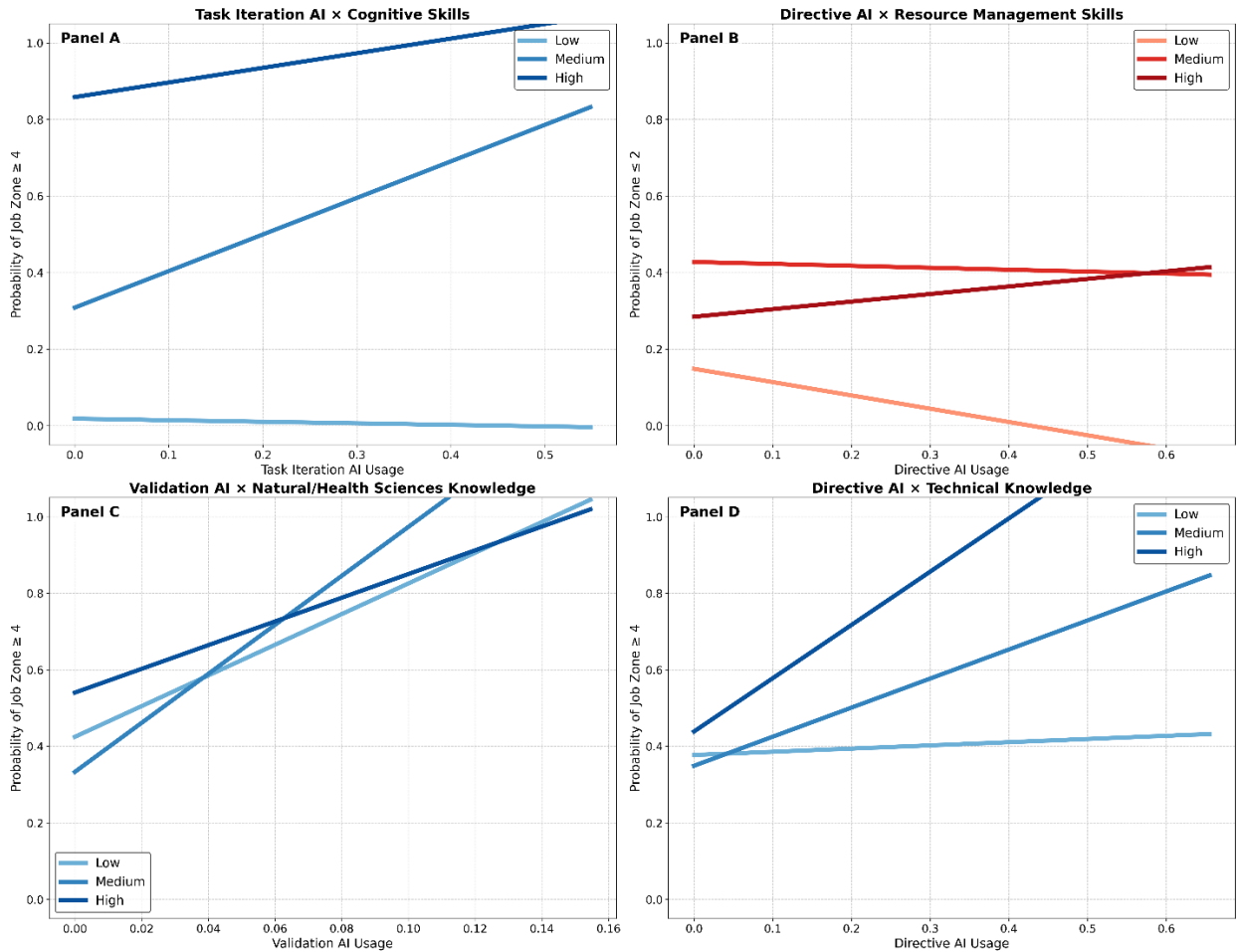

Note: Plots show predicted probabilities from ordinal logistic regression for different levels (Low/Medium/High) of KSA component scores, illustrating significant interaction effects from Table 1.

### 5.4. Comparative Predictive Power

To test H4, we compared nested models with and without AI usage metrics. The inclusion of AI usage metrics significantly improved model fit compared to a model with only skill dimensions. The McFadden's Pseudo R² increased from 0.48 to 0.54, indicating that AI usage patterns explain a meaningful proportion of variance in job zone placement beyond what can be accounted for by skill dimensions alone. To test H4, we compared nested models. Table 2 presents the fit statistics, and Figure 4 highlights the improvement in McFadden's Pseudo R².

**Table 2: Comparison of Model Fit Statistics for Job Zone Prediction Models**

| Performance Metric | Base Model (KSA Only) | Full Model (With Interactions) | Improvement (Full - Base) |
|---|---|---|---|
| McFadden's Pseudo R-squared | 0.48 | 0.54 | +12.1% |
| Log-Likelihood (Fitted) | -408.68 | -401.29 | +7.39 |
| AIC | 855.36 | 840.58 | -14.78 |
| BIC | 937.08 | 922.30 | -14.78 |

Note: Comparison between KSA Model and Full Model (with KSA × AI interaction terms). Lower AIC/BIC values and higher R²/Log-Likelihood indicate better fit, supporting H4.

**Figure 4: Improvement in Model Fit (McFadden's Pseudo R²) with Sequential Inclusion of AI Usage Metrics and KSA-AI Interaction Terms**

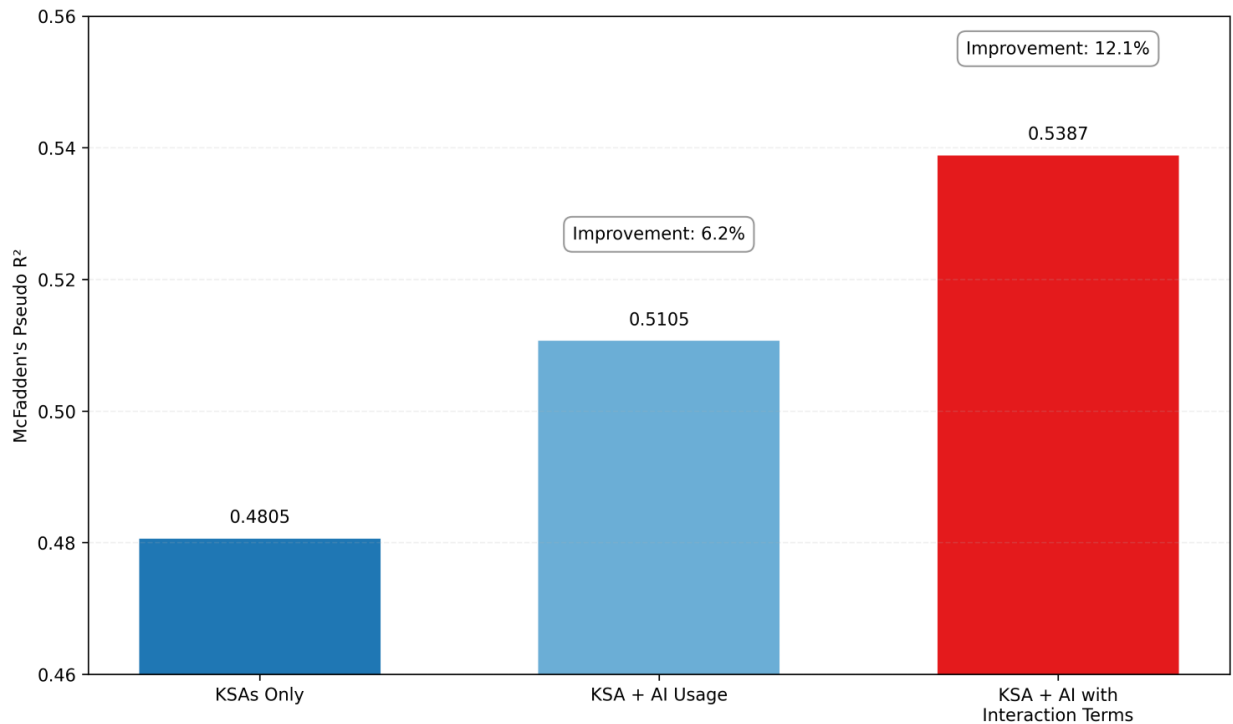

Note: Comparison of model fit with and without AI usage and interaction terms

This finding confirms that the 2ACT framework captures a distinct dimension of occupational classification that is not fully reflected in traditional skill measures. AI usage patterns represent an emerging axis of differentiation in the labor market that shapes career advancement opportunities independently of established skill requirements.

6. Discussion

Our analysis provides novel insights into how 2ACT shapes advancement opportunities, highlighting both anticipated and unexpected relationships that challenge conventional narratives about AI's impact on occupational mobility. The 2ACT framework demonstrates significant explanatory power beyond traditional skill-based classification models, suggesting that AI usage patterns constitute an emerging dimension of occupational stratification with profound implications for career pathways.

## 6.1. Theoretical Contributions
### 6.1.1. Differentiating AI Impact Across Occupational Hierarchies

Our results empirically confirm the missing-middle pattern (Daugherty & Wilson, 2018): automation-focused interactions (directive, feedback loop) are strongly associated with lower job-zone placement whereas augmentative interactions (task iteration, thinking fraction) predict movement into higher zones. This asymmetric impact aligns with task-biased technological-change theory (Autor et al., 2003; Acemoglu & Autor, 2011) and supports Autor's (2015) claim that substitution technologies cluster in middle-skill work while complementary technologies enhance high-skill roles (Brynjolfsson & Mitchell, 2017). Figure 3 sharpens this point. Only workers who combine high cognitive KSAs with iterative AI leverage experience appreciable upward mobility, suggesting that AI amplifies, rather than equalizes, existing skill differentials. This amplification effect has significant implications for wage inequality and indicates the need for examining intra-occupational patterns where heterogeneous AI-usage may further widen skill premiums. Furthermore, the observed strong negative relationship between feedback loop interactions and job zone placement extends Jarrahi's (2018) and Jarrahi et al.'s (2023) observations that feedback-intensive AI applications often appear in routinized service contexts where they structure and constrain worker autonomy. Feedback loop AI implementations may create algorithmic cages (Shestakofsky, 2017) that constrain skill development and advancement opportunities.

### 6.1.2. Skill-AI Interactions as Mobility Bridges

Our most consequential findings emerge in the interaction effects between KSA components and AI usage patterns. The strong positive interaction between task iteration and cognitive/analytical skills represents the most promising skill bridge identified in our analysis. This interaction effect substantially exceeds the already positive main effect of cognitive/analytical skills, suggesting that the combination of these skills with iterative AI collaboration creates particularly strong career advancement potential.

This finding extends research on skill complementarity in important ways. While Deming (2017) identified the growing importance of cognitive skills combined with social skills, our results suggest that in an AI-transformed economy, cognitive skills combined with specific AI collaboration patterns may constitute a distinct form of complementarity that facilitates vertical mobility across job zones (Blair et al., 2020). This is particularly notable for those engaged in iterative refinements. This extends Alabdulkareem et al.'s (2018) concept of skill bridges by identifying not just clusters of complementary skills but specific skill-technology interaction patterns that enhance mobility.

The significant positive interaction between validation AI and natural/health sciences knowledge reveals another potential mobility pathway aligned with Davenport and Kirby's (2016) stepping in pattern where humans maintain oversight of AI systems. This creates what Frank et al. (2019) described as complementary intelligence that appears particularly valued in higher job zones, characterized by human domain expertise combined with AI analytical capabilities.

Equally revealing are the negative interaction effects, particularly directive AI × resource management skills and learning AI × resource management skills. These negative interactions indicate skill-AI combinations that predict lower job zone placement, representing mobility barriers rather than bridges. Routine-intensive management tasks are particularly susceptible to AI substitution (Tolan et al., 2021), and our results identify specific skill-AI combinations that appear especially vulnerable to downward pressure.

The counterintuitive positive interaction between directive AI and engineering/physical sciences knowledge reveals a more complex pattern than simple substitution versus complementarity narratives would suggest. In domains requiring specialized technical knowledge, directive AI may complement rather than substitute for human expertise. This is a pattern Webb (2020) identified in specialized engineering fields where AI systems execute technical processes while humans maintain domain oversight, an emerging human in the loop phenomenon.

### 6.1.3. Reconceptualizing Occupational Classification

Our findings contribute to broader dialogue about determinants of occupational classification by demonstrating that AI usage patterns represent a distinct dimension not fully captured by traditional skill measures. The significant improvement in model fit with the addition of AI metrics indicates that the 2ACT framework identifies a previously unexamined axis of differentiation in the labor market. AI similarity reveals an emerging domain with the potential to identify advancement pathways invisible within conventional occupational classification systems, necessitating a fundamental reconceptualization of how we understand career mobility in an AI-transformed economy.

## 6.2. Stakeholder Implications
### 6.2.1. Strategic Career Development for Individuals

For workers seeking advancement in an AI-transformed economy, our findings suggest specific strategies for skill investment and AI interaction. The strong positive relationship for cognitive/analytical skills combined with the significant interaction effect with task iteration AI indicates that developing both analytical capabilities and collaborative fluency with iterative AI systems creates particularly valuable advancement potential. Conversely, workers in occupations characterized by directive and feedback loop AI interactions should prioritize developing skills that facilitate transitions to occupations with different AI usage profiles, particularly those emphasizing augmentative patterns.

### 6.2.2. Curriculum Design for Educational Institutions

Educational institutions should incorporate specific AI collaboration modalities into curriculum design rather than treating "AI skills" as a monolithic category. Programs should develop students' capacity for iterative and validation-oriented AI interactions that show positive associations with career advancement while teaching critical awareness of how directive and feedback loop interactions might constrain development. Most importantly, institutions should focus on cultivating the specific skill-AI combinations, notable cognitive skills with task iteration AI, that function as effective skill bridges for upward mobility.

### 6.2.3. Policy Interventions for Workforce Development

Policymakers should reconceptualize career pathway programs to incorporate AI usage patterns as dimensions of occupational similarity and mobility potential (Escobari et al., 2019). Traditional frameworks based primarily on industry or educational credentials may miss emerging mobility opportunities created by similar AI usage patterns across dissimilar occupations. Targeted upskilling initiatives should focus on developing the specific skill-AI combinations identified as effective bridges for career advancement, particularly for workers in occupations characterized by automation-focused AI usage patterns that predict lower job zone placement. Extending the work of Tolan et al. (2021), these recommendations target the development of human capabilities that complement rather than compete with AI benchmarks.

### 6.2.4. Implementation Approaches for Organizations

Organizations implementing AI systems should consider the career implications of different interaction patterns. The negative association between directive/feedback loop interactions and job zone placement suggests these implementation approaches may inadvertently create advancement barriers. Conversely, implementation approaches emphasizing task iteration and validation may enhance career development opportunities. Organizations concerned with internal mobility should prioritize augmentative AI implementations, particularly for roles where retention and advancement are strategic priorities.

## 6.3. Limitations and Future Directions

Several limitations suggest important directions for future research. Our cross-sectional design prevents causal inference regarding relationships between AI usage patterns and job zone placement. Longitudinal research tracking changes in AI usage patterns and corresponding shifts in job zone classifications would provide stronger evidence regarding causal mechanisms.

Our occupation-level analysis may mask within-occupation heterogeneity in AI usage patterns. Task-level analysis could provide more granular insights into how AI-accentuated task transformation creates differentiated advancement opportunities within occupational categories.

Finally, our AI usage metrics represent current patterns of human-AI collaboration, which will likely evolve as capabilities advance. Future research tracking changes in AI usage patterns over time would provide valuable insights into the dynamics of career mobility in an increasingly AI-integrated economy.

## 6.4. Conclusion

By differentiating between distinct AI usage patterns and examining their relationship with job zone placement, our study provides a nuanced understanding of how 2ACT are shaped by the interplay between AI usage patterns and KSAs. Our identification of specific skill bridges through interaction effects between KSA components and AI usage patterns provides a foundation for developing targeted advancement strategies in today's AI-transformed economy.

Most importantly, our results suggest that 2ACT creates new forms of occupational connectivity based on similar AI usage patterns rather than traditional skill similarities. AI usage similarity represents a novel dimension of career mobility networks that may reveal previously invisible advancement pathways across conventional occupational boundaries. Understanding these emerging bridging mechanisms will be fundamental to developing effective career strategies, educational curricula, and workforce policies that leverage rather than resist the transformative potential of AI technologies.

**Data Availability**

The data supporting this study are available from publicly accessible repositories but are not provided with the article due to licensing restrictions. The Anthropic/EconomicIndex dataset is hosted on Hugging Face (https://huggingface.co/datasets/Anthropic/EconomicIndex), and the O*NET files are available via the O*NET Resource Center (https://www.onetcenter.org/database.html#individual-files). Interested parties may obtain the data directly from these sources under their respective terms and conditions.

Peterson, N. G., Mumford, M. D., Borman, W. C., Jeanneret, P. R., Fleishman, E. A., Levin, K. Y., Campion, M. A., Mayfield, M. S., Morgeson, F. P., Pearlman, K., Gowing, M. K., Lancaster, A. R., Silver, M. B., & Dye, D. M. (2001). Understanding work using the Occupational Information Network (O*NET): Implications for practice and research. *Personnel Psychology, 54*(2), 451-492. https://psycnet.apa.org/doi/10.1111/j.1744-6570.2001.tb00100.x

Raisch, S., & Krakowski, S. (2021). Artificial intelligence and management: The automation-augmentation paradox. *Academy of Management Review, 46*(1), 192-210. https://psycnet.apa.org/doi/10.5465/amr.2018.0072

Rothwell, J. (2015). *Defining skilled technical work*. National Academies of Sciences, Engineering, and Medicine. https://dx.doi.org/10.2139/ssrn.2709141

Shestakofsky, B. (2017). Working algorithms: Software automation and the future of work. *Work and Occupations, 44*(4), 376-423. https://doi.org/10.1177/0730888417726119

Sullivan, S. E., & Arthur, M. B. (2006). The evolution of the boundaryless career concept: Examining physical and psychological mobility. *Journal of Vocational Behavior, 69*(1), 19-29. https://doi.org/10.1016/j.jvb.2005.09.001

Sullivan, S. E., & Baruch, Y. (2009). Advances in career theory and research: A critical review and agenda for future exploration. *Journal of Management, 35*(6), 1542-1571. https://psycnet.apa.org/doi/10.1177/0149206309350082

Tibshirani, R. (1996). Regression shrinkage and selection via the lasso. *Journal of the Royal Statistical Society: Series B (Methodological), 58*(1), 267-288. https://www.jstor.org/stable/2346178

Tolan, S., Pesole, A., Martínez-Plumed, F., Fernández-Macías, E., Hernández-Orallo, J., & Gómez, E. (2021). Measuring the occupational impact of AI: Tasks, cognitive abilities and AI benchmarks. *Journal of Artificial Intelligence Research, 71*, 191-236. https://doi.org/10.1613/jair.1.12647

Vaccaro, M., Almaatouq, A., & Malone, T. (2024). When combinations of humans and AI are useful: A systematic review and meta-analysis. *Nature Human Behaviour*, *8*, 2293–2303. https://doi.org/10.1038/s41562-024-02024-1

Vendramin, N., Nardelli, G., & Ipsen, C. (2021). Task-Technology Fit Theory: An approach for mitigating technostress. In C. Ipsen & L. L. Saling (Eds.), *A handbook of theories on designing alignment between people and the office environment* (pp. 39–53). Routledge.

Webb, M. (2020). *The impact of artificial intelligence on the labor market* [Working paper]. Stanford University, Department of Economics. https://www.michaelwebb.co/webb_ai.pdf.
25